\documentclass{svjour3}                     
\smartqed  
\usepackage{graphicx}
\begin{document}

\title{Instruments of RT-2 Experiment onboard CORONAS-PHOTON
and their test and evaluation V: Onboard software, Data Structure, Telemetry and Telecommand
\thanks{This work was made possible in part from a grant from Indian Space Research Organization
(ISRO). The whole-hearted support from G. Madhavan Nair, Ex-Chairman, ISRO, who initiated the 
RT-2 project, is gratefully acknowledged.
}
}

\titlerunning{RT-2 experiment: Onboard software }   

\author{S. Sreekumar \and P. Vinod \and Essy Samuel \and  J. P. Malkar \and A. R. Rao \and M. K. Hingar 
\and V. P. Madhav \and D. Debnath \and T. B. Kotoch \and Anuj Nandi \and S. Shaheda Begum
\and Sandip K. Chakrabarti}

\authorrunning{Sreekumar et al. } 

\institute{S. Sreekumar, 
P. Vinod, Essy Samuel \at
Vikram Sarabhai Space Centre, VRC, Thiruvananthapuram 695022\\
              \email{sreekumar\_s@vssc.gov.in}\\
	\and 
         J. P. Malkar, A. R. Rao, M. K. Hingar, V. P. Madhav \at
Tata Institute of Fundamental Research, Homi Bhabha Road, Colaba, Mumbai 400005\\
           \and
            D. Debnath, T. B. Kotoch, A. Nandi$^+$ \at
Indian Centre for Space Physics, 43 Chalantika, Garia Station Rd., Kolkata 700084\\
Tel.: +91-33-24366003\\
              Fax: +91-33-24622153 Ext. 28\\
              \email{dipak@csp.res.in; tilak@csp.res.in; anuj@csp.res.in}   
($+$: Posted at ICSP by Space Science Division, ISRO Head Quarters, Bangalore)
      \and
     S. Shaheda Begum\at
Radio Astronomy Centre, NCRA-TIFR, Ooty 643001 \\
      \and
Sandip K. Chakrabarti \at
              S.N. Bose National Centre for Basic Sciences, JD Block, Salt Lake, Kolkata 700097 \\
(Also at Indian Centre for Space Physics, 43 Chalantika, Garia Station Rd., Kolkata 700084)\\
              Tel.: +91-33-23355706\\
              Fax.: +91-33-23353477\\
              \email{chakraba@bose.res.in}           
}

\date{Received: date / Accepted: date}

\maketitle

\begin{abstract}

The onboard software and data communication in the RT-2 Experiment onboard the Coronas-Photon 
satellite is organized in a hierarchical way to effectively handle and communicate asynchronous 
data generated by the X-ray detectors. A flexible data handling system is organized in the
X-ray detector packages themselves and the processing electronic device, namely RT-2/E, 
has the necessary intelligence to communicate with the 3 scientific payloads by issuing commands 
and receiving data. It has direct interfacing with the Satellite systems and issues commands to 
the detectors and processes the detector data before sending to the satellite systems. The
onboard software is configured with several novel features like a) device independent communication
scheme, b) loss-less data compression and c) Digital Signal Processor. Functionality of the onboard 
software along with the data structure, command structure, complex processing scheme etc. are 
discussed in this paper. 


\end{abstract}

\section{Introduction}
\label{intro}

RT-2 Experiment onboard the Coronas-Photon satellite (Kotov et. al. 2008) consists of 3 scientific 
and 1 processing electronic payloads (Nandi et al. 2009a). The processing electronic device, namely, 
RT-2/E communicates with the scientific payloads and the ground stations 
through SSRNI (System of Collection and Registration of Scientific Informations 
or SCRSI in English) and the Control and Communication unit (BUS-FM) of the satellite. The three 
scientific payloads are RT-2/S \& RT-2/G (both Phoswich scintillating detectors of NaI(Tl)/CsI(Na)
crystals) and RT-2/CZT (solid-state imaging detector). 
In Debnath et al. (2009), Kotoch et al. (2009), Nandi et al. (2009), we described the technical
aspects of three scientific payloads, their functionality and different imaging techniques that 
are designed and implemented in the RT-2/CZT payload.
In Sarkar et al. (2009), we mainly discussed the effect of the cosmic diffuse high energy
X-ray background on all the three detectors through Monte-Carlo simulations
as implemented in GEANT-4 toolkit.

The RT-2 Experiment covers the energy range of 15 to 150 keV extendable up to 1 MeV. All 
three  payloads have different fields of view ranging from 4$^\circ$x 4$^\circ$ (RT-2/S), 
6$^\circ$x 6$^\circ$ (RT-2/G) and 6$'$ --  6$^\circ$ (RT-2/CZT). In order 
to view the sky (Sun) in the low energy $\gamma$-ray range, all the three detector 
systems are placed outside the hermetically sealed module of the satellite. 
The three detectors are mounted with instrument axis parallel to the Sun pointing axis 
of the satellite. On the other hand, RT-2/E along with other processing systems of the satellite 
are placed inside the hermetically sealed chamber of the satellite.

In the present paper, we will concentrate on the onboard software and the overall 
functionality of RT-2/E. The schematic diagram of RT-2 system is shown in Figure 1.

\begin{figure}[h]
\includegraphics[height=2.5in]{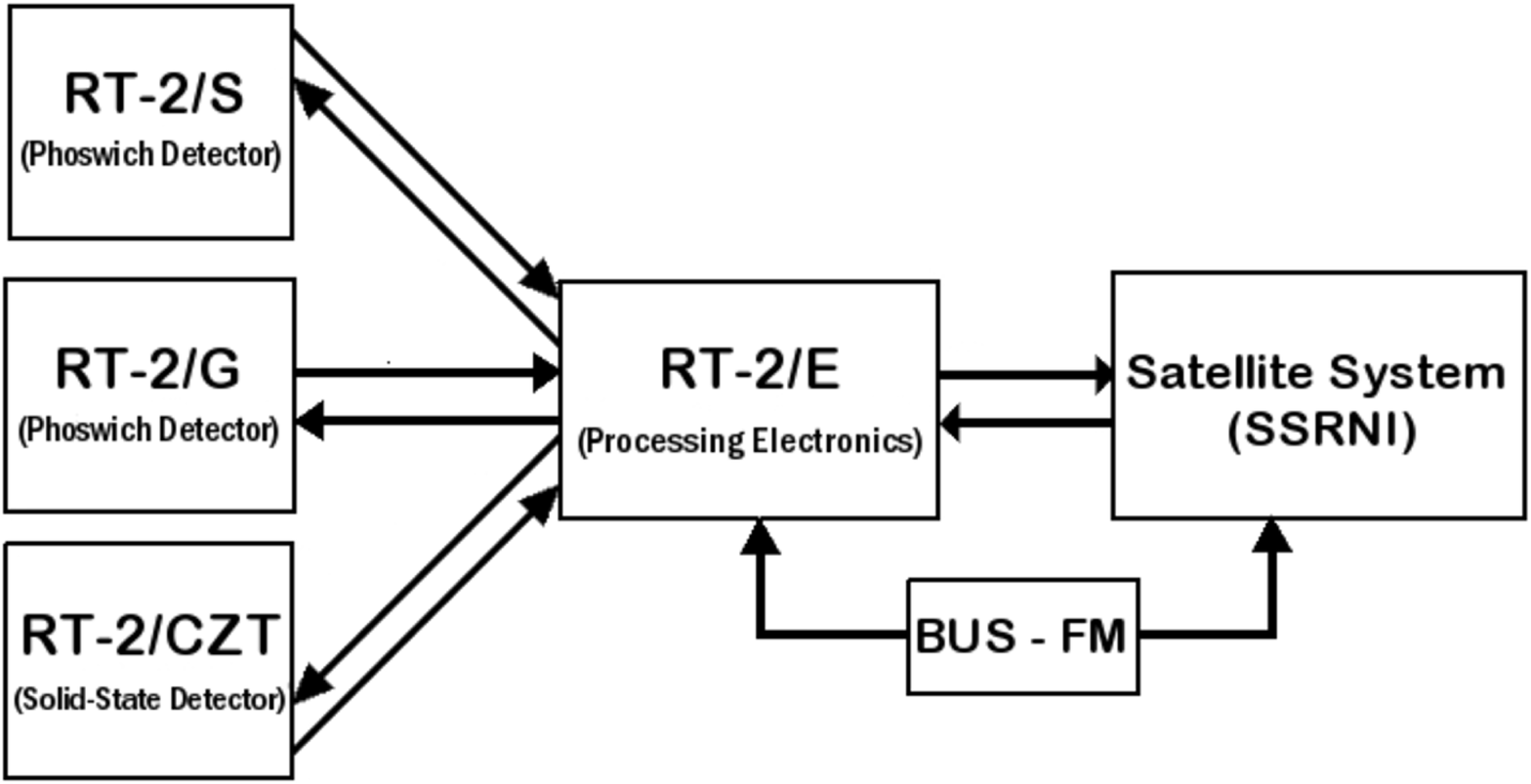}
\caption{Schematic diagram of RT-2 system.}
\label{}
\end{figure}


\section{Description of the processing electronics (RT-2/E)}

RT-2/E is the main processing electronic device of the RT-2 Experiment and it
acts as an interface between the detector electronics and the satellite system. 
The unit decodes the telecommand appropriately and transmits the detector data to the ground 
through the satellite telemetry, which also involves the functions of compressing and making 
packets of the data from the detectors. The control logic system of the device is FPGA (Field 
Programmable Gate Array), which carries all the logical operations in the RT-2/E. A Digital 
Signal Processor ADSP2101 is used in the system for data processing. The RT-2/E electronics 
also consists of input and output buffers for detectors, satellite interface, electronics for 
power interface, memory interface. The schematic block diagram of RT-2/E device is shown in 
Figure 2. A continuous power supply to the RT-2 system is provided by the Control and 
communication unit (BUS-FM) of the satellite. Power supply is provided by a DC current source 
with the Voltage of 27$^{+7}_{-3}$ Volt without a midpoint. The maximum power provided for the 
RT-2 system is 32 Watt.

RT-2/E weighs 8.56 kg and the power consumption of the device is 3.78 Watt, $\le$ 10 kg and  
$\le$ 5 Watt, respectively, as per design requirement. RT-2/E is operable in the temperature 
range of -10$^\circ$C to +40$^\circ$C and it is placed inside a hermetically sealed chamber of 
the satellite.
  
\begin{figure}[h]
\includegraphics[height=3.4in]{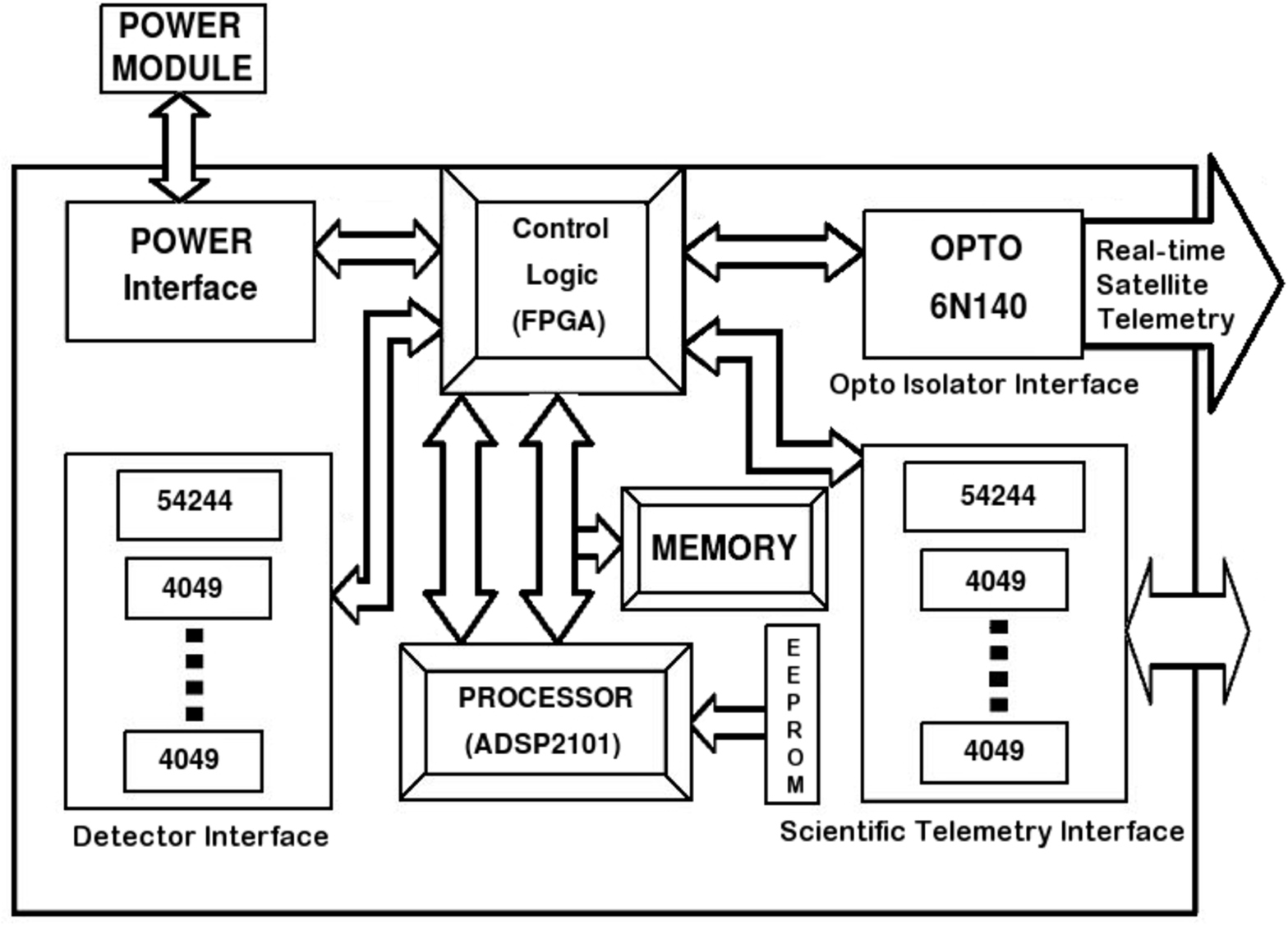}
\caption{Schematic block diagram of the processing electronic device RT-2/E.}
\label{}
\end{figure}

\section{Basic functions of RT-2/E}

In the power-on mode, all detectors detect X-rays and package them in a `page' in the detector box 
itself, in a mode called the `normal' mode, which can be changed to `test' mode by command. Every 
second, RT-2/E sends a `second' signal to the detectors and the data is sent to RT-2/E while the 
storing is done in a separate `page' (these two pages toggle every second). RT-2/E processes the 
detector data and sends to the satellite. In RT-2/E, the data is made into packets with the 
processing intervals known as frames in whole (a number of packets as one frame) and each frame is 
divided into blocks. These blocks are compressed and are written in the memory of RT-2/E, in the 
form of packets. This telemetry data is subsequently sent to the satellite.
Telecommands, which are up-linked to the satellite, include commands to adjust high voltage (HV), 
low level discriminator (LLD) value, Channel boundary change, `mode' change and so on. All these 
commands are decoded within RT-2/E and are passed on to the respective detectors. 

Another important task of the RT-2/E is to execute the pulse commands (SWITCH OFF/ON) that are 
available form the Control and communication unit (BUS-FM) of the satellite. The control commands 
(pulse commands) of the RT-2 system are discussed later on.

\section{Role of RT-2 Onboard Software}

The telemetry resources available for the RT-2 Experiment is 32 bits of satellite telemetry,
read every 4 seconds and transmitted semi-continuously to the ground and 10 Mbytes of onboard memory, 
transmitted once or twice daily to the ground. The scientific requirement is to have coarse spectral
and timing information of the X-ray data on a continuous basis and higher time 
resolution information during transient events. Further, it is also desirable to
have faster read out of information for ground testing as well as 
for trouble shooting purposes. The onboard software needs to be parameterized so that a
sufficient flexibility is available to change the working of the software based on 
a 16-bit data command and 14 `pulse commands'. For time synchronization, calibration
time information is also available from the satellite subsystem SSRNI.
This multiple and complex demands are realized in the following way:

\begin{itemize}
\item[{\large $\bullet$}] The asynchronous data from the X-ray detectors are packaged 
in the detector blocks themselves. This packaging is organized in two modes: a) a time
tagged event mode to get 0.3 ms time resolution (to cater to any special needs like high 
time resolution study of some celestial objects and also for debugging the detector software)
and b) a spectral and timing mode which has a time resolution better than what is
scientifically required. Spectrum and image every second, and timing information every 10 
millisecond, packaged every second are deemed to be sufficient to satisfy all the
requirements. Time-stamping is done in the detector block using a local clock, and the
data is interrogated and taken precisely every second, and this precision is established
by taking this information from SSRNI. The detector performance and software are
controlled by a few data commands, designed as a subset of the 16-bit satellite 
data command.

\item[{\large $\bullet$}]  The basic detector data are packaged and kept in a memory
in RT-2/E, to be transmitted to the satellite memory with its own protocol. 
This re-packaging is done based on modes and while transmitting, data are compressed
using a loss-less compression code.

\item[{\large $\bullet$}] The information from the detector is codified (8 bits per payload)
and sent to the satellite telemetry every 4 seconds to have a basic diagnostic of the
working of the experiment.

\item[{\large $\bullet$}]  The modes of operation is done based on ground commands as well as
onboard processing. The onboard processing caters to a) flare detection b) memory 
availability c) satellite position (high and low background regions) and solar visibility.

\item[{\large $\bullet$}] Facility is also kept to change the complete onboard software.
\end{itemize}

The data collected from the detectors is first taken into input buffers of RT-2/E memory and is 
accumulated in the accumulation buffers during a processing interval. On every processing interval 
boundary or on every processing mode change, the frame of data is compressed and made into packets 
by the onboard software. The algorithm used for loss-less compression of RT-2/E telemetry data is 
the CCSDS (Consultative Committee for Space Data Systems) recommended Rice algorithm 
(Yeh et al. 1991; Rice et. al. 1993). The compressed 
data is then sent to the satellite and then subsequently to the ground. Also, all the telecommands 
coming from the ground are decoded and appropriate commands are sent to the corresponding detectors.

The onboard flare detection logic is enabled every second. Satellite telemetry data is prepared 
by the software every second. Also, all the required functions are carried out on receiving the 
ON/OFF commands from the satellite. 

Memory management scheme is included in the software, which does the work of managing the memory
in RT-2/E. Further, CZT detectors initialization data is sent to CZT from the RT-2/E memory, on 
receiving CZT initialization command from ground. Watchdog timer updating logic is incorporated in 
the software, which will enable the hardware to reset the digital signal processor in case of 
software hang up. 

There is an additional feature of downloading the contents of the program memory (onboard software) 
for verification. This can be enabled based on ground command.
The onboard software is developed in the assembly language of ADSP2101 processor.

\section{Modes of Operation}

Based on the diverse constraints for using the satellite memory as well as requirement of the 
scientific interest, various processing modes are defined for RT-2/E.

\subsection{Bad Mode (Mode Id = 0, 100 sec/frame)}

The bad mode runs when the satellite enters into high flux region (South Atlantic
Anomaly - SAA, North and South polar regions), based on signals from the satellite 
called `GOOD' and `BAD'. When RT-2/E is in this mode, the onboard software lowers the high 
voltage of Photomultiplier Tubes (PMT) of the Phoswich detectors and makes the HV of CZT 
detectors to zero and only the frame header data is transmitted but not the detector data. 
When coming out of this mode i.e., to the good state, the previous modes are started afresh 
and the high voltage is set for the PMTs and CZT. The switch over to good mode from bad mode 
can be delayed by ground command as multiples of 64 seconds. In the bad mode, the data (header 
data) is sent every 100 seconds.

\subsection{Test Mode (Mode Id = 1, 1 sec/frame)}

The software enters this mode, when the test mode data from the detectors are received. There will 
be house keeping data (VCO) and event data from the detectors, in this mode. The most significant 
bit of VCO data, if set, takes the software to this mode. The raw data from the detectors is sent to 
the satellite as soon as they arrive to RT-2/E, i.e., every second without any compression.
There are also command based facilities to send a limited number of CZT events and
also to send only the CMOS data.

\subsection{Debug Mode (Mode Id = 2, 1 sec/frame)}

In this mode, the spectral data from detector units is sent to the satellite every second as soon 
as they arrive to RT-2/E. The purpose of this mode is debugging during the initial verification 
phase as well as in the case of any later malfunctions of either packages (payloads). RT-2/E is 
switched to the debug mode on commands from ground.

When RT-2/E is in the debug mode, there is an option to get the timing data alone from the 
detectors.

\subsection {Solar Quiet Mode - SQM (Mode Id = 4, 100 sec/frame)}

This is the primary accumulation mode since the Sun is quiet in hard X-rays most of the time. In 
this mode, spectrum is obtained for every 100 second and count rates for every second in RT-2/S 
and G. Similarly, the spectra and the images in every 100 second and the count rates in every 
second are obtained for CZT detectors while only images are obtained for CMOS in every 100 second.  

\subsection{Solar Flare Mode - SFM (Mode Id = 3, 10 sec/frame)}

The major science requirement for this experiment is the availability of high temporal and spectral 
resolution data during solar flares. Since such flares occur randomly, the onboard software has a 
built-in mechanism for checking the current count rate against the present thresholds to detect the 
flares. The flare search is carried out at every second. The logic of flare detection will be 
discussed later. The data packaging is unaffected for RT-2/CZT during flares.

In this mode, data frame structure is identical to the solar quiet mode except that both the time 
resolutions are reduced by a factor of 10, i.e., in this mode, the count rates are 
stored at every 0.1 second and spectra are stored at every 10 second.

Normally, the data accumulation is done in the quiet mode. However, as soon as the flare trigger 
occurs, the current frame of the quiet mode is filled to the next multiple of 10 s. The quiet 
mode data till the detection of flare is made into packets and a new frame is started in the 
flare mode. After trigger, the data is accumulated in this mode for next 10 second i.e., one 
frame. At the end of the frame, again flare threshold is checked and if the count rate is still 
more than that, then this mode will continue for another frame. Otherwise, the data accumulation 
will revert back to the quiet mode. Flare mode data frames are stored in the same address stream 
of data as the quiet mode.

\subsection {Shadow Mode (Mode Id = 4, 100 sec/frame)}

Shadow mode is activated when the Sun is out of the detector field of view i.e., 
during NIGHT and at the time of solar occultation. In this mode, flare detection is disabled. The 
processing of data in the shadow mode is similar to the solar quiet mode (SQM). The detector data 
are stored for every 100 seconds in blocks of 64 words. While transmitting, they are compressed 
and transmitted in packets of 60 words (59 words of data and one packet header). Also, for each 
type of data a separate packet called frame header is created.
 
For example, in the normal mode for RT-2/S and G, after decompression at ground, one frame data 
consists of 57 packets data (in RT-2/CZT 218 packets). Out of these 57 packets of data, first packet 
data is called frame header. Rest of the 56 packets contain scientific data. First word of each data 
packet is called packet header, it signifies current packet number out of total packets in the frame. 
The number of packets after decompression (60 words each) will be number of blocks before compression 
* 64/59 + 1 and these values are estimated and given in Table 1.

\begin{table}[h]
\renewcommand{\arraystretch}{1.3}
\caption{Modes \& packetisation of RT-2/S, RT-2/G and RT-2/CZT payload data} 
\label{FirstTable}
\centering
\begin{tabular}{|c|c|c|c|c|c|}
\hline
{\bf Processing Mode} & {\bf Description} & \multicolumn{2}{c|}{\bf Compressed}& \multicolumn{2}{c|}{\bf Decompressed}\\
\cline{3-6}
	      &		         & {\bf S/G } & {\bf CZT } & {\bf S/G } & {\bf CZT } \\
\hline
Bad Mode & Frame Header, every 100 sec & - & - & 1 & 1 \\
\hline
Shadow mode & Spectrum every 100 sec & 51 & 200 & 57 & 218 \\
\hline
Test mode & Event data every sec & 1-230 & 1-198 & 2-251 & 2-216 \\
	  & Software download & - & 64 & - & 71 \\
\hline
Debug mode & Spectrum every sec & 51 & 200 & 57 & 218 \\
	   & Timing alone & 13 & 19 & 16 & 22 \\
\hline
Solar flare mode & Spectrum every 10 sec & 51 & - & 57 & - \\
\hline
Solar quiet mode & Spectrum 100 sec & 51 & 200 & 57 & 218 \\
\hline
\end{tabular}
\end{table}


\subsection {Mode Selection Logic}

The mode of operation of the experiment depends on the input from the satellite system (GOOD/BAD 
and LIGHT/SHADOW), data commands to the detector (to decide Test/Normal mode), data commands to 
the processor (to decide Normal/Debug mode), and onboard analysis (to decide memory availability 
and flare detection). Again, the flare detection logic can be fine-tuned 
using several data commands. Further, any of the data commands can be given in a time-tagged mode.

\begin{table}[h]
\renewcommand{\arraystretch}{1.3}
\caption{Good data condition for RT-2 payloads} 
\label{SecondTable}
\centering
\begin{tabular}{|c|c|c|c|c|c|}
\hline
{\bf GOOD/BAD} & {\bf LIGHT/SHADOW} & {\bf Detector Mode} & {\bf Processor Mode} &  {\bf Flare detected} & {\bf Output Mode}   \\
\hline
Bad & X & X & X & X & Bad \\
\hline
Good & X & Test & X & X  & Test \\
\hline
Good & Shadow & Normal & Normal & X & SQM \\
\hline
Good & X & Normal & Debug & X & Debug \\
\hline
Good & Light & Normal & Normal & No & SQM \\
\hline
Good & Light & Normal & Normal & Yes & SFM \\
\hline 
\end{tabular}
\end{table}


The selection logic for the Mode of operation is given in Table 2, which is valid when the 
available memory is $>$50\%. The letter `X' in the table implies that the condition of that 
particular column is ignored to determine the Mode. The ``BAD'' condition is given the highest
priority, when no scientific data is transmitted but only the frame header, i.e., one packet 
containing vital health parameters, is transmitted every 100 s. When the available memory is 
$<$25\%, the output is deemed to be in the Bad mode. The detector `Test Mode' is given the next 
priority, when the output is in the `Test Mode'. This mode is enabled to directly get the detector 
data at higher time resolution for trouble-shooting purposes and hence very rarely used. 
Similarly, the `Debug Mode' of the processor is used to directly transmit the detector data for 
trouble shooting purposes.

The normal mode of operation is the SQM. When the satellite is in the `Light' region, the flare 
trigger is activated to take data transmission mode to SFM. Flare detection, however, is disabled 
when the available memory is below 50\%. Sufficient care is taken such that the mode transitions
do not cause any break in the data transmission (for example, when the data storing goes from SFM 
to SQM, it is done at an integer multiple of 10 seconds). Extensive and elaborate information is 
given in the header (see Table 20) so that the onboard logic which caused the mode transitions
could be clearly understood.


\section {Detector Data Formats}

\subsection {Data format for the RT-2/S and RT-2/G detectors}

In the detector device, the asynchronous data from the NaI (Tl) / CsI (Na) crystals 
(Debnath et al. 2009) are stored in the memory. This data is transmitted to RT-2/E every second, 
based on a command from RT-2/E. Thus the 
asynchronous data from the detectors are sent in a synchronized manner for further processing. 

The data storage occurs in two modes: i) test mode, where every event is time-tagged correct to 
$0.3$ ms and ii) normal mode, where all the spectral data is accumulated and 
eight channel count rates are stored every 10 ms and sent in every second. 
The basic data (for each registered X-ray event) consists of  pulse height (PH),
pulse shape (PS), amplifier from which pulse height is determined (gain 1 - G1
or gain 2 - G2), and the time of registration of events. The detector
box also creates histograms of pulse height depending on whether they are from NaI (Tl) or
CsI (Na) (based on whether the pulse shape, PS, is less than a predetermined value, $PS_{cut}$.
Hence the highest time resolution possible from the experiment is: a) 0.3 ms in the event mode and 
b) 10 ms for 8 channel counters and c) 1 second for full spectrum. Apart from the detector data, the 
house-keeping (HK) information of the detectors is also sent to RT-2/E by encoding the information 
through a Voltage Controlled Oscillator (VCO). The spectral data is packaged in the detector device 
and dispatched to RT-2/E in the following format.

\subsubsection {Test mode}

Each detector `event' is characterized by 2 words and event data structure is given in Table 3.

\begin{table}[h]
\renewcommand{\arraystretch}{1.3}
\caption{Event data structure of RT-2/S \& RT-2/G.} 
\label{ThirdTable}
\centering
\begin{tabular}{|c|c|c|c|}
\hline
{\bf D31-D20} & {\bf D19-D13} & {\bf D12} & {\bf D11-D0}   \\
\hline
Time & Pulse Shape & G1 or G2 & Energy  \\
\hline 
\end{tabular}
\end{table}

Maximum events that can be stored in memory are 7360 events.
This data block is stored in a memory and sent to RT-2/E, whenever 1 second command is 
received. The channel number is incremented every second.

\subsubsection {Normal mode}

\begin{enumerate}

\item {Header block (1 word): VCO data (2 bytes)}
\item Spectrum block (2432 words):  G1 Na (PH): 1K words; + G1 Cs (PH): 1 K words; + G2: 256 words; + 
width: 128 words.
\item Timing blocks (800 words): 100 Timing blocks X 8 counters X 1 word (counters in each block will 
count for 10 ms).
\item Counter block (8 words): 8 counters (1 word each)

\end{enumerate}

The house-keeping data of Phoswich detectors (RT-2/S \& RT-2/G) are multiplexed through VCO and
stored in 8 channel format. The VCO data structure (2 byte) is given in Table 4.

\begin{table}[h]
\renewcommand{\arraystretch}{1.3}
\caption{VCO data structure of RT-2/S and G payloads.} 
\label{ForthTable}
\centering
\begin{tabular}{|c|c|c|c|}
\hline
{\bf D15} & {\bf D14} & {\bf D13-D11} & {\bf D10-D0}   \\
\hline
Mode ID & Corona & VCO Channel No. & VCO Counts  \\
\hline 
\end{tabular}
\end{table}

The detail descriptions of eight VCO channels (HK parameter) is given in Table 5 (NC means 
Not Connected). One of these channel parameters is sent to RT-2/E through VCO every second.

\begin{table}[h]
\renewcommand{\arraystretch}{1.3}
\caption{HK parameters of RT-2/S and G payloads.} 
\label{First Table}
\centering
\begin{tabular}{|c|c|c|}
\hline
{\bf Channel No.} & {\bf Description} & {\bf Voltage level}   \\
\hline
0 & Supply Voltage & + 5V $\pm$ 0.5V  \\
\hline 
1 & Thermistor & + 1.5V - +5.0V  \\
\hline 
2 & Supply Voltage & + 5V $\pm$ 0.5V  \\
\hline 
3 & NC & -  \\
\hline
4 & HV Feedback & + 1.5V - +5.0V  \\
\hline 
5 & HV Reference & + 1.5V - +5.0V  \\
\hline 
6 & LLD Voltage & + 1.5V - +5.0V  \\
\hline 
7 & NC & -  \\
\hline 
\end{tabular}
\end{table}

Note: VCO frequency (2 kHz min. to 20 kHz max.) is corresponding to 1 to 10 Volt variation.

\subsection {Data format for the RT-2/CZT detector}

In the detector block, the asynchronous data from the CZT detectors (Kotoch et al. 2009) are stored 
in memory. This data is transmitted to RT-2/E every second, based on 1 second command from RT-2/E. 
Thus the asynchronous data from the detectors are sent in a synchronized manner for further 
processing. 

The data storage occurs in two modes: i) test mode where every event is time-tagged correct to 
0.3 ms and ii) normal mode, where all the spectral and image data are accumulated and 
count rates are stored every 10 ms and sent in every second. Hence the highest time resolution 
possible from the RT-2/CZT is: a) 0.3 ms in the event mode and 
b) 10 ms for 12 channel counters and c) 1 second for full spectrum and image. 
Apart from the detector data, the House-Keeping (HK) information of the detectors is also sent to 
RT-2/E by encoding the information through an ADC. 

RT-2/CZT is set to command mode by ground command before issuing commands to individual CZT modules. 
For CZT, the ADC value corresponding to each pixel is multiplied by a gain factor 1.xx, where 'xx' 
varies from 0 - 0.25 in steps of 0.001 (to correct the errors in the energy range and count ratio of 
pixels). Then a constant value (CZT constant can be commanded from ground) is subtracted from this. 
This constant is the same for all pixels. Further, an offset value is added to this. Offset is an 
8 bit number and can be different for each pixel.
 
$\bullet$ Pixel value (CZT) = (ADC O/P value) * gain - CZT constant + offset

Here, the `CZT constant' is subtracted to account for the sign of the offset value. The optimum 
gain and offset values for each pixel is stored in the gain-offset table within the CZT memory. 
The ADC value, after applying these operations is used for image and spectrum computation in 
the normal mode. In the test mode, upper 10 bits of the actual ADC value (12 bits) are used in 
the event data.

For CMOS detector, 512 x 1024 pixel values are read periodically and a CMOS constant (which can be 
changed by ground command) is subtracted from the ADC value corresponding to each pixel. 

$\bullet$ {Pixel value (CMOS) = (ADC O/P value) - CMOS constant} 

An upper threshold value is also defined (ground command) and if the pixel value crosses this 
threshold, the value is zeroed, considering it as a bad pixel. For 1-bit image, if the value 
crosses a threshold (defined separately for each line), the value is 1, else 0. 
 
Bit image of the CMOS detector is made by applying logical OR to the values of four adjacent 
(2x2) pixels. This is done for 512 x 512 pixels, starting from a start line. Start line and line 
thresholds can also be changed by command. This results in a 256 x 256 1-bit image. The sum of all 
vertical pixels and horizontal pixels are computed separately (before `OR'ing) as $V_{sum}$ and 
$H_{sum}$ respectively. 

In CMOS, frames (each scanned image) are obtained depending on the frame interval provided. Frame 
interval is the time set to get a frame from the detector, which is selectable from ground. For 
example, if the frame interval is set for 4 seconds, a frame is obtained every 4 seconds. Frames are 
integrated in the detector itself during this interval. In the test mode, all the frames received 
in every 32 seconds are integrated within RT-2/CZT and the integrated data is sent to RT-2/E in 
32 second (data of 8k pixels in each second). The possible frame intervals in normal mode and test 
mode are given in Table 6.

\begin{table}[h]
\renewcommand{\arraystretch}{1.3}
\caption{Frame Intervals in Normal and test modes of RT-2/CZT (CMOS detector).} 
\label{}
\centering
\begin{tabular}{|c|c|}
\hline
                &       \\
\hline
{\bf Normal Mode} & {\bf Test Mode}   \\
\hline
 1 & 1  \\
\hline 
 2 & 2  \\
\hline 
 4 & 4  \\
\hline 
 5 & 8  \\
\hline
 10 & 16  \\
\hline
 20 & 32  \\
\hline
 25 & -  \\
\hline
 50 & -  \\
\hline
 100 & -  \\
\hline
\end{tabular}
\end{table}

{\bf CMOS Calibration}:

Calibration of CMOS pixels can be done in the test mode of RT-2/CZT. Data of N consecutive image 
frames are averaged and these frames are taken, where N is 32 divided by the frame interval, 
defined in the previous table. The calibration result is written in 256 $V_{sum}$ locations, two 
values 
in each word. The calibration result is identified by `1' in the location 8182 of RT-2/CZT data.
Addition of an offset value to the calibrated results can also be done by command in normal or test 
mode.

The contents of RT-2/CZT EEPROM can be sent as data to RT-2/E in test mode, on receiving the command 
0x4803. The data from RT-2/CZT is sent to RT-2/E every second in the following format.

\subsubsection {Test mode}

Each CZT `event' is characterized by 2 words and event data structure is given in Table 7.
This data is stored in the memory and sent to RT-2/E, every second. Maximum events that can be 
stored in memory are 4032 events.

\begin{table}[h]
\renewcommand{\arraystretch}{1.3}
\caption{Event data structure of RT-2/CZT detector} 
\label{First Table}
\centering
\begin{tabular}{|c|c|c|c|}
\hline
{\bf D31-D20} & {\bf D19-D10} & {\bf D9-D2} & {\bf D1-D0}   \\
\hline
Time & ADC value of Detector data & Pixel ID & Detector No. (0-2 for 3 CZTs) \\
\hline 
\end{tabular}
\end{table}

\subsubsection {Normal mode}

The normal mode data format for CZT (3 modules) and CMOS detector is given below:

{\bf Data of CZT}:

\begin{enumerate}

\item Image block (3072 words): 1 K words per CZT (4 channel X 256 pixels X 1 word).
\item Spectrum block (1536 words):  512 words per CZT
\item Timing blocks (1200 words): 3 CZT detector X 100 timing words X 4 channels X 1 word (counters 
in each block will count for 10 ms).
\item Counter block (24 words): 12 counters (2 words each)
\item VCO block (1 word): 2 bytes of ADC data

\item Special words (8 words) : Satellite telemetry word, temperature, Command sent, Data read against 
command, event number , CMOS line number, Calibration result identification word and Calibration 
status 

\end{enumerate}

{\bf Data of CMOS}:

\begin{enumerate}
\item Image block (4096 words): 256 x 256 pixels
\item Sum (512 words): Vertical sum (256 words) + Horizontal sum (256 words)
\end{enumerate}

The  house-keeping (HK) data of CZT-CMOS detector (RT-2/CZT) are multiplexed through an ADC and HK
parameters are stored in 8 channel format. The ADC data structure is given in Table 8.

\begin{table}[h]
\renewcommand{\arraystretch}{1.3}
\caption{ADC data structure of CZT-CMOS detector} 
\label{}
\centering
\begin{tabular}{|c|c|c|c|}
\hline
{\bf D15} & {\bf D14} & {\bf D13-D11} & {\bf D10-D0}   \\
\hline
Mode ID (0: Normal, 1: Test) & 1: EEPROM, 0: Detector & ADC Channel No. & ADC Counts  \\
\hline 
\end{tabular}
\end{table}

The detail description of eight ADC channels (HK parameter) is given in Table 9. One of these channel
parameters is sent to RT-2/E through ADC in every second.

\begin{table}[h]
\renewcommand{\arraystretch}{1.3}
\caption{HK Parameters of RT-2/CZT payload} 
\label{}
\centering
\begin{tabular}{|c|c|c|}
\hline
{\bf Channel No.} & {\bf Description} & {\bf Voltage level}   \\
\hline
0 & Supply Voltage & + 5V $\pm$ 0.5V  \\
\hline 
1 & Thermistor & + 2.5V - +5.0V  \\
\hline 
2 & NC & -  \\
\hline 
3 & NC & -  \\
\hline
4 & HV Control &  0.0/3.3V  \\
\hline 
5 & CZT Supply (DVDD) & + 3.6V  \\
\hline 
6 & CMOS Supply & + 5.0V  \\
\hline 
7 & FPGA core Supply (Vcca)& + 2.5V  \\
\hline 
\end{tabular}
\end{table}

Note: In `command' mode of CZT, Data read against CZT detector commands comes in the special words.





\section{Accumulation of Detector Data}

Data from the detectors are first fed to the input buffers, where the spectral data / event data 
of the detectors are buffered. There are two separate buffers for each detector. Data is written 
alternately into each buffer. FPGA writes the detector data to one of the buffers and the 
processor reads the data from the other buffer. The allocation of all three detector data in normal 
mode is discussed in the following sections.

\subsection {Input buffer of RT-2/S \& RT-2/G:}

A total of 3328 words of memory space are allocated for the RT-2/S and RT-2/G detectors data in 
the input buffers. Data structure in the input buffer is given in Table 10.

\begin{table}[h]
\renewcommand{\arraystretch}{1.3}
\caption{Data structure of the input buffers of RT-2/S \& RT-2/G: } 
\label{}
\centering
\begin{tabular}{|c|c|c|}
\hline
{\bf Address} & {\bf Data Type} & {\bf Data (words)}   \\
\hline
0-1023 & NaI Spectrum & 1024  \\
\hline 
1024-2047 & CsI Spectrum & 1024  \\
\hline 
2048-2847 & Timing & 800  \\
\hline 
2936-2943 & Total Counts & 8  \\
\hline
2944-3071 & PSD Spectrum &  128  \\
\hline 
3072-3327 & G2 Spectrum & 256  \\
\hline 
\end{tabular}
\end{table}

\subsection {Accumulation buffer of RT-2/S \& RT-2/G}

The data from the input buffers is accumulated in an accumulation buffer. A total of 3248 words of 
memory space are allocated for the RT-2/S and RT-2/G detectors data in the accumulation buffers. The 
data allocation in this accumulation buffer is given in Table 11.

\begin{table}[h]
\renewcommand{\arraystretch}{1.3}
\caption{Data structure of the accumulation buffer of RT-2/S and RT-2/G} 
\label{11thTable}
\centering
\begin{tabular}{|c|c|c|}
\hline
{\bf Address} & {\bf Data Type} & {\bf Data (words)}   \\
\hline
0-1023 & NaI Spectrum & 1024  \\
\hline 
1024-2047 & CsI Spectrum & 1024  \\
\hline 
2048-2303 & G2 Spectrum & 256  \\
\hline 
2304-2431 & PSD Spectrum & 128  \\
\hline
2432-3231 & Timing & 800  \\
\hline 
3232-3247 & Counts & 16  \\
\hline 
\end{tabular}
\end{table}

The 8 extra words in the accumulation buffer are to accommodate deeper (2 words each) accumulation 
of counters.

\subsection {Input buffer of RT-2/CZT:}

RT-2/CZT is an imaging device. It has two different type of detectors, namely CZT and CMOS. The image, 
spectrum and timing data allocation of CZT of RT-2/CZT in input buffers is described in Table 12.

\begin{table}[h]
\renewcommand{\arraystretch}{1.3}
\caption{Data structure of CZT in RT-2/CZT payload.} 
\label{First Table}
\centering
\begin{tabular}{|c|c|c|}
\hline
{\bf Address} & {\bf Data Type} & {\bf Data (words)}   \\
\hline
0-4095 & Image & 4096  \\
\hline 
4096-6143 & Spectrum & 2048  \\
\hline 
6144-7343 & Timing & 1200  \\
\hline 
7744-7767 & Counts & 24  \\
\hline
8176 & CZT Satellite Telemetry word & 1  \\
\hline 
8177 & Temperature & 1  \\
\hline 
8178 & CZT Command & 1  \\
\hline 
8179 & Data read against Command & 1  \\
\hline 
8180 & Event Number & 1  \\
\hline 
8181 & CMOS line Number & 1  \\
\hline 
8182 & 1: Threshold calibration result in vertical sum location & 1  \\
\hline 
8183 & 0: CMOS calibration & 1  \\
\hline 
\end{tabular}
\end{table}


In the input buffer, 8 words are assigned for 'some' special informations (see Table 12) for 
RT-2/CZT. Description of special words are given below:

Temperature word:


        Bits 7-0      :    Temperature,
        Bits 9-8      :    CZT number,
        Bits 13-10  :    0xF valid data,
        Bits 15-14  :    base address of RT-2/CZT EEPROM \\

Event number:

       Bit 15 : CZT3 ON,
       Bit 14 : CZT2 ON,
       Bit 13 : CZT1 ON,
       Bits 12-0 : (number of events)/64 \\

Line number:

       Bits 7-0   : CMOS line number,
       Bits 15-8 : frame interval \\

CMOS detector has only image information. A total of 4608 words of memory space are allocated for 
CMOS detector data in the input buffers. The image data allocation of CMOS detector in input buffer 
is given in Table 13.

\begin{table}[h]
\renewcommand{\arraystretch}{1.3}
\caption{Data structure of CMOS detector data} 
\label{First Table}
\centering
\begin{tabular}{|c|c|c|}
\hline
{\bf Address} & {\bf Data Type} & {\bf Data (words)}   \\
\hline
8192 - 12287 & Image & 4096  \\
\hline
12288 - 12799 & Sum & 512 \\
\hline
\end{tabular}
\end{table}

\subsection {Accumulation buffer of RT-2/CZT:}

A total of 5832 words of memory space are allocated for CZT detector data in the accumulation 
buffers. Data allocation structure in accumulation buffer is given in Table 14.

\begin{table}[h]
\renewcommand{\arraystretch}{1.3}
\caption{Data structure in accumulation buffer for CZT of the RT-2/CZT detector.} 
\label{First Table}
\centering
\begin{tabular}{|c|c|c|}
\hline
{\bf Address} & {\bf Data Type} & {\bf Data (words)}   \\
\hline
0 - 3071 & Image & 3072  \\
\hline 
3072 - 4607 & Spectrum & 1536  \\
\hline 
4608 - 5807 & Timing & 1200  \\
\hline 
5808 - 5831 & Counts & 24  \\
\hline
\end{tabular}
\end{table}

Data allocation of CMOS detector in accumulation buffer is identical as of input buffer and data 
structure is given in Table 15.

\begin{table}[h]
\renewcommand{\arraystretch}{1.3}
\caption{Data structure of CMOS of RT-2/CMOS } 
\label{First Table}
\centering
\begin{tabular}{|c|c|c|}
\hline
{\bf Address} & {\bf Data Type} & {\bf Data (words)}   \\
\hline
8192 - 12287 & Image & 4096  \\
\hline
12288 - 12799 & Sum & 512 \\
\hline
\end{tabular}
\end{table}

During test mode, data in input buffer of RT-2/CZT is read by RT-2/E for telemetry and in normal 
mode, data in accumulation buffer for telemetry. The allocation of RT-2/CZT data in RT-2/E input 
buffer in test mode is given in Table 16.

\begin{table}[h]
\renewcommand{\arraystretch}{1.3}
\caption{Structure of the allocation of RT-2/CZT data in RT-2/E.  } 
\label{First Table}
\centering
\begin{tabular}{|c|c|c|}
\hline
{\bf Address} & {\bf Data Type} & {\bf Data (words)}   \\
\hline
0 - 8063 & CZT Event reports & 8064  \\
\hline
8176 - 8183 & CZT Special words & 8 \\
\hline
0 - 8192 & EEPROM data & 8192 \\
\hline
8192 - 12287 & CMOS test data & 4096 \\
\hline
12288 - 12543 & CMOS calibration results & 256 \\
\hline
\end{tabular}
\end{table}

\section{Data Compression Scheme}

RT-2 telemetry data is subjected to loss-less data compression, for effective bandwidth as well 
as memory utilization. Each frame of data is divided into different blocks, each block of size 
64 words. Each block is compressed, made into packets and then written to RT-2/E memory. CCSDS 
recommended Rice Algorithm (Yeh et al. 1991; Rice et. al. 1993) is used for compression. 
There are various compression options in the algorithm, the best-suited option 
for each block is decided onboard and the selected option is used 
for compressing the block of data. If compression is not achieved for the data block, actual data 
will be sent. The compressed data is made into packets and is written in the memory. Total data 
size in each frame, number of compressed block and block size is summarized below:

$\bullet$ Number of telemetry data per frame: 3248 words for RT-2/S \& RT-2/G,
                                    12800 words (including 7760 words for CZT, 
                                    and 4608 for CMOS) for RT-2/CZT,
                                    4096 Program memory words.

$\bullet$ Number of blocks to be compressed: 51 for RT-2/S \& RT-2/G data,
			     	   200 for CZT data,
				   64 for Program memory data.

$\bullet$ Block size: 64 \\

Each block data contains a block header of one word followed by the compressed data and is written 
in the packet. Flow chart to compress data is given below:

$\bullet$ Get blocks of data 

$\bullet$ Set first sample as the reference sample
 
$\bullet$ Check the difference between successive samples 

$\bullet$ Select the best compression option for the block

$\bullet$ Compress the differences using the best option 

$\bullet$ Packet the compressed data 

\subsection {Compression Options:}

The block header word (16 bits) of each block has the following information:

Bits 15-8: Compression option,
Bits 7-0: Block number. \\

The difference of adjacent samples in each block is computed and the difference values are coded 
using the best option given in Table 17.

\begin{table}[h]
\renewcommand{\arraystretch}{1.3}
\caption{Compression options of RT-2/E telemetry data.} 
\label{First Table}
\centering
\begin{tabular}{|c|c|}
\hline
{\bf Option No.} & {\bf Option}   \\
\hline
0 & 16:0  \\
\hline
1 & 15:1  \\
\hline
2 & 14:2  \\
\hline
4 & 12:4  \\
\hline
5 & zero block  \\
\hline
6 & No compression  \\
\hline
\end{tabular}
\end{table}

{\bf No compression option}

When this option is chosen, the data consists of block header followed by 64 data samples of the 
block.

{\bf Zero block}

When all the samples in the block are the same, all the difference values will be zeroes. In this 
case, this option is chosen. Here, for each block, compressed data consists of one block header 
followed by the first sample of the block.

{\bf 16:0 option}

Here, difference value as such will be coded using Rice code.

{\bf 15:1 option}

In this option, each difference value/2 is encoded using Rice code. The LSB of the difference is 
appended as such. Thus, for a block, compressed data consists of three parts - block header, Rice 
codes of all the differences and followed by the LSBs (Least Significant Bit) of all differences. 
The LSBs are packed into words and written, starting from a fresh word. In this option, there will 
be 4 LSB words, following the encoded data.  

{\bf 14:2 option}

In this option, each difference value/4 is encoded using Rice code. The two LSBs of the difference 
is appended as such. In this option, there will be 8 LSB words, following the block header and 
encoded data.

{\bf 12:4 option}

In this option, each difference value/16 is encoded using Rice code. The four LSBs of the difference 
is appended as such. In this option, there will be 16 LSB words, following the block header and 
encoded data.

The fundamental sequence of Rice code is given in Table 18.

\begin{table}[h]
\renewcommand{\arraystretch}{1.3}
\caption{The fundamental sequence in the Rice code.} 
\label{First Table}
\centering
\begin{tabular}{|c|c|}
\hline
{\bf Value} & {\bf Code}   \\
\hline
0 & 1   \\
\hline
1 & 01   \\
\hline
-1 & 001   \\
\hline
2 & 0001   \\
\hline
-2 & 00001   \\
\hline
. & .   \\
\hline
. & .   \\
\hline
\end{tabular}
\end{table}

\section{Packeting Scheme}

The detector data, after compression, is written in RT-2/E memory in the form of packets. Each 
frame data results a number of packets, having 60 words (960 bits) each. Out of these packets, 
first packet is frame header and the remaining packets are data packets. Frame header is written 
into memory as such without compression. All data packets (from second packet) consist of 
compressed data. First word of each data packet is the packet header. 


The packet header (Table 19) will have the packet number and the location where the next block of 
compressed data starts in the packet. If there is no new block data in the packet, this number will 
be zero. The packet header location in the first packet (frame header) will always be 0x0101.

\begin{table}[h]
\renewcommand{\arraystretch}{1.3}
\caption{Packet header in RT-2/E memory.} 
\label{First Table}
\centering
\begin{tabular}{|c|c|}
\hline
{\bf Bit No.} & {\bf Description}   \\
\hline
15 - 8 & Packet number   \\
\hline
7 - 0 & Starting location of first block of compressed data \\
\hline
\end{tabular}
\end{table}


The description of the 60 words (1 packet) in the frame header is given in the Table 20,
21 and 22.

\begin{table}[h]
\renewcommand{\arraystretch}{1.3}
\caption{Description of the frame header in RT-2/E memory.} 
\label{First Table}
\centering
\begin{tabular}{|c|c|c|}
\hline
{\bf Word No.} & {\bf Description} & {\bf Remarks}  \\
\hline
0  & Packet header - 0x0101 & - \\
\hline
1  & Detector ID   & 0: RT-2/S, 1:RT-2/G, 2:RT-2/CZT, 3: EEPROM packet \\
\hline
2  & Frame No.   & No. of current frame \\
\hline
3  & Time duration of frame   & sec \\
\hline
4  & Time LSB  & Time in sec. \\
\hline
5  & Time MSB  & Time in sec. \\
\hline
6  & GPS Time LSB (interface19)  & From LSB: 10 bits: millisec; 6 bits: sec \\
\hline
7  & GPS Time MSB (interface19)  & 6 bits: minutes; 5 bits: hours; 5 bits: days \\
\hline
8  & GPS Time LSB (interface20)  & From LSB: 10 bits: millisec; 6 bits: sec \\
\hline
9  & GPS Time MSB (interface20)  & 6 bits: minutes; 5 bits: hours; 5 bits: days \\
\hline
10  & Read packet number - Telemetry1 (TM1)  & Number of packets read from TM1 memory \\
\hline
  & Read packet number - Telemetry2 (TM2)  & Number of packets read from TM2 memory \\
\hline
11  & Write packet number - Telemetry1 (TM1)  & Number of packets written in TM1 memory \\
\hline
  & Write packet number - Telemetry2 (TM2)  & Number of packets written in TM2 memory \\
\hline
12  & Processing mode ID  & 0: Bad mode, 1: Test mode, 2: Debug mode, 3: SFM, 4: SQM \\
\hline
13  & Bit14 - second select  & 0: sec in interface19, 1: sec in interface20 \\
\hline
    & Bit13 - good veto  & 1: veto good, 0: otherwise \\
\hline
    & Bit12 - bad veto  & 1: veto bad, 0: otherwise \\
\hline
    & Bit11 - light veto  & 1: veto light, 0: otherwise \\
\hline
    & Bit10 - shadow veto  & 1: veto shadow, 0: otherwise \\
\hline
    & Bit9 - millisecond select & 0: ms in interface 19, 1: ms in interface 20 \\
\hline
    & Bit8 - GPS valid & 1: valid GPS in interface20, 0: invalid GPS in interface 20 \\
\hline
    & Bit7 - second boundary flag for interface20 & 1:second boundary \\
\hline
    & Bit6 - command flag for interface20 & 1 when received, 0 when not received \\
\hline
    & Bit5 - 4 & unused bits \\
\hline
    & Bit3 - solar mode & 1: flare mode, 0: quiet mode \\
\hline
    & Bit2 - processor mode & 1: debug mode, 0: normal mode \\
\hline
    & Bit1 - corona & 1: corona\\
\hline
    & Bit0 - detector mode & 1: test mode, 0: normal mode\\
\hline
14   & Status port1* & *See Table 21  \\
\hline
15   & Satellite Telemetry LSB & Telemetry channels of RT-2/G and RT-2/S  \\
\hline
16   & Satellite Telemetry MSB & Telemetry channels of RT-2/E and RT-2/CZT  \\
\hline
17   & Data Count & Count of data received from detector input buffer  \\
\hline
18   & Flare Threshold & Value of the threshold set (0-255) (Fth)\\
\hline
      & 		 & if (C1+C2+C5)/32 (sum count rate) $\ge$ Fth, then flare flag set\\
\hline
19    & High Voltage command & Current command given for HV \\
\hline
20-27    & VCO data (8 words) & VCO channels of the detectors \\
\hline
28    & Memory availability**  & **See Table 22 \\
\hline
29    & Ground command & Last ground command given \\
\hline
30    & Bit15 (CZT and CMOS data flag) & 1: CMOS data alone in CZT test mode \\
\hline
       &	 &  0: CZT+CMOS data in CZT test mode \\
\hline
    & Bit14 (Timing flag) & 1: timing alone in debug mode, 0: all data in debug mode \\
\hline
    & Bit13-0 (Number of event data) & Valid only in test mode \\
\hline
    & {\bf RT-2/S \& RT-2/G} &  \\
\hline
31-46    & Total counts (16) & C1-C8 counter values, 2 words for each counter \\
\hline
47-59    & Ground commands (13) & 13 Ground commands given lastly \\
\hline
    & {\bf RT-2/CZT} &  \\
\hline
31-54    & Total counts (24) & C1-C12 counter values, 2 words for each counter \\
\hline
55-59    & CZT data (5) & Word 55: Temperature, Word 56: command \\
\hline
       &               & Word 57: data read against command \\
\hline
         &              & Word 58: event number, Word 59: CMOS line number \\
\hline
\end{tabular}
\end{table}

\begin{table}[h]
\renewcommand{\arraystretch}{1.3}
\caption{*Status Port1} 
\label{First Table}
\centering
\begin{tabular}{|c|c|c|}
\hline
{\bf Bit No.} & {\bf Description} & {\bf Remarks}  \\
\hline
15-7  & Unused & - \\
\hline
6  & Memory area & 0: section1, 1: section2 \\
\hline
5  & GPS valid (interface19) & 0: valid GPS, 1: invalid GPS \\
\hline
4  & Input buffer select & 0: buffer1, 1: buffer2 \\
\hline
3  & Second flag (interface19) &  1: second boundary \\
\hline
2  & Command (interface19) &  1: command received, 0: otherwise \\
\hline
1  & Good/Bad &  0: Good, 1: Bad \\
\hline
0  & Light/Shadow &  0: Light, 1: Shadow \\
\hline
\end{tabular}
\end{table}

\begin{table}[h]
\renewcommand{\arraystretch}{1.3}
\caption{**Memory availability} 
\label{First Table}
\centering
\begin{tabular}{|c|c|c|}
\hline
{\bf Bit No.} & {\bf Availability} & {\bf Processing modes supported}  \\
\hline
0  & $\ge$50\% & All modes \\
\hline
1  & 50\% & All except SFM \\
\hline
2  & 25\% & Bad mode \\
\hline
3  & Nil & No data \\
\hline
\end{tabular}
\end{table}

\section{Satellite Interface}

A sophisticated interface scheme is provided for RT-2/E with the Coronas-Photon satellite.
Satellite interface scheme includes pulse commands and power ON/OFF for RT-2 payloads, telecommand
interface line, scientific telemetry for down link detector data and `real-time' satellite telemetry
to have instruments health informations in every 4 seconds. Details of satellite interface with 
RT-2/E and specific requirements are discussed bellow:
 
1. Power and Pulse Commands (PC) for ON/OFF (14 contact commands):

These commands are required for power ON/OFF of the instruments and selection of GOOD/BAD operating
region in the satellite orbit. All command informations are summarized in the Table 23.

\begin{table}[h]
\renewcommand{\arraystretch}{1.3}
\caption{Power and Pulse commands of RT-2 operation} 
\label{First Table}
\centering
\begin{tabular}{|c|c|c|}
\hline
{\bf Contact Identification} & {\bf Contact} & {\bf Contact type}  \\
\hline
PC124 & Power ON (RT-2) & Executed in BUS-FM \\
\hline
PC125 & Power OFF (RT-2) & Executed in BUS-FM \\
\hline
PC130 & Switch ON (RT-2/S) & Pulse dry contact  \\
\hline
PC131 & Switch OFF (RT-2/S) & Pulse dry contact \\
\hline
PC132 & Switch ON (RT-2/G) & Pulse dry contact \\
\hline
PC133 & Switch OFF (RT-2/G) & Pulse dry contact \\
\hline
PC134 & Switch ON (RT-2/CZT) & Pulse dry contact \\
\hline
PC135 & Switch OFF (RT-2/CZT) & Pulse dry contact \\
\hline
PC136 & Change Operational Mode & Pulse dry contact \\
\hline
PC137 & RESET RT-2 & Pulse dry contact \\
\hline
PC147 & LIGHT & Continuous dry contact \\
\hline
PC148 & SHADOW & Continuous dry contact \\
\hline
PC149 & BAD & Continuous dry contact \\
\hline
PC150 & GOOD & Continuous dry contact \\
\hline
\end{tabular}
\end{table}

2. Data commands (16 bits): 

To have the redundancy in telecommand, two separate telemetry interfaces 19 and 20 are provided for 
each telecommand chain i.e. 5 lines X 2 parallel (`AND'ed). These are the five parallel lines, which 
are named as `one second command', `time sync', `command sync', `clock', `serial data' for 
telecommands transmitted in two independent channels.

3. Scientific telemetry: 

Scientific telemetry is the transmission of the scientific (spectral/event) data from 
the detectors. Detector data are packetized in RT-2/E depending on the processing mode. This 
scientific telemetry is done in two separate telemetry channels TM1 (for RT-2/S and RT-2/G) and 
TM2 (for RT-2/CZT). 

4. Satellite telemetry (32 bits): 

Satellite telemetry is a `real-time' telemetry scheme, which sends the data (health parameters)
in each 4 second. Satellite telemetry is done in 32 bits (total) - 8 bits for three detectors (S, 
G and CZT) and for the electronics device (RT-2/E), which are sent every 4 second. In these 32 bits, 
15 bits are allotted for the three 5 bit counters (S, G and CZT), 3 bits for 3 VCO bytes converted 
into bits and 8 bits for the parameters (status of memory, orbit etc.) in RT-2/E. Four digital 
channels, each of 8-bit wide are used to sent the informations to the satellite telemetry by RT-2/E. 
In Table 24, all informations are summarized. 

\begin{table}[h]
\renewcommand{\arraystretch}{1.3}
\caption{Description of Satellite Telemetry bits (32 bits)} 
\label{First Table}
\centering
\begin{tabular}{|c|c|c|}
\hline
{\bf Bit No.} & {\bf Description} & {\bf Remarks}  \\
\hline
              &      {\bf RT-2/S}             &                 \\
\hline
0       &   +5 V         & 1 = +5V, 0 = 0V \\
\hline
1       &   Corona ON/OFF        & 1 = Corona ON, 0 = Corona OFF \\
\hline
2       &   HV Feedback       & 1 = HV ON, 0 = HV OFF \\
\hline
3 - 7    & 5 bit counter  &   (C1+C2+C5)/32 \\
\hline
         &    {\bf RT-2/G}			&			\\
\hline
8       &   +5 V         & 1 = +5V, 0 = 0V \\
\hline
9       &   Corona ON/OFF        & 1 = Corona ON, 0 = Corona OFF \\
\hline
10       &   HV Feedback       & 1 = HV ON, 0 = HV OFF \\
\hline
11 - 15    & 5 bit counter  &   (C1+C2+C5)/32 \\
\hline
         &    {\bf RT-2/CZT}	&			\\
\hline
16       &   +5 V         & 1 = +5V, 0 = 0V \\
\hline
17       &   Command mode: CZT status      & 1 = Command sent to Detector, 0 = Command not sent \\
         &   Event mode: Sum of Vsum     & 1 = sum of Vsum $>$ threshold, 0 = sum of Vsum $<$ threshold \\
         &   EEPROM data read mode   & Bit 1 of data in address 8176 of EEPROM* \\
\hline
18      &   HV Feedback       & 1 = HV ON, 0 = HV OFF \\
\hline
19 - 23    & 5 bit counter  &   (Channel2 count of CZT1)/32 \\
	    & In EEPROM read mode   &  bits 3-7 of data in address 8176 of EEPROM* \\
\hline
         &    {\bf RT-2/E}			&			\\
\hline
24       &   LIGHT/SHADOW        & 1 = SHADOW, 0 = LIGHT \\
\hline
25      &   GOOD/BAD        & 1 = BAD, 0 = GOOD \\
\hline
26       &   Flare       & 1 = ON, 0 = OFF \\
\hline
27    & +5 V  &  1 = +5V, 0 = 0V \\
\hline
28 - 29    & Memory availability (TM1)  &  0 = $>$50\%, 1 = 50\%, 2 = 25\%, 3 = NIL\\
\hline
30 - 31    & Memory availability (TM2)  &  0 = $>$50\%, 1 = 50\%, 2 = 25\%, 3 = NIL\\
\hline

\end{tabular}
*These slots can be used to verify EEPROM lock/unlock.
\end{table}

\section{Ground Commands}

There will be certain situations to adjust some parameters like the high voltage to the PMT, the 
LLD value and channel boundary value. These can be done from ground by sending commands to the 
satellite. Also, the data transfer and mode selection are done by the ground commands. All detector 
commands are passed through the RT-2/E processing device. Details of detector commands are 
summarized in respective papers (Debnath et al. 2009; Kotoch et al. 2009). RT-2/E commands are 
summarized in Table 25.
These commands include processor mode selection (SFM, SQM, Debug, Test etc.) with different condition, 
Flare selection logic for S and G, reseting the RAM, BAD to GOOD time set duration etc.

\begin{table}[h]
\renewcommand{\arraystretch}{1.3}
\caption{Ground commands of RT-2/E} 
\label{First Table}
\centering
\begin{tabular}{|c|c|c|}
\hline
{\bf Command Type} & {\bf Command} & {\bf Description}  \\
\hline
Normal mode (RT-2/S) &  0x80xx    &  Set RT-2/S in normal mode with flare threshold of xx  \\
\hline
Debug mode (RT-2/S) &  0x8800    &  Set RT-2/S in debug mode to get all the data from it  \\
\hline
Debug mode with timing alone (RT-2/S) &  0x8801    &  Set RT-2/S in debug mode to get timing data alone from it  \\
\hline
Test mode with `n' events (RT-2/S) &  0x89xx    &  Set RT-2/S in test mode, d6-d0: no. of events/64  \\
\hline
Normal mode (RT-2/G) &  0xA0xx    &  Set RT-2/G in normal mode with flare threshold of xx  \\
\hline
Debug mode (RT-2/G) &  0xA800    &  Set RT-2/G in debug mode to get all the data from it  \\
\hline
Debug mode with timing alone (RT-2/G) &  0xA801    &  Set RT-2/G in debug mode to get timing data alone from it  \\
\hline
Test mode with `n' events (RT-2/G) &  0xA9xx    &  Set RT-2/G in test mode, d6-d0: no. of events/64  \\
\hline
Normal mode (RT-2/CZT) &  0xC000    &  Set RT-2/CZT in normal mode  \\
\hline
Debug mode (RT-2/CZT) &  0xC800    &  Set RT-2/CZT in debug mode to get all the data from it  \\
\hline
Debug mode with timing alone (RT-2/CZT) &  0xC801    &  Set RT-2/CZT in debug mode to get timing data alone from it  \\
\hline
Test mode, only CMOS &  0xC9FF    &  Set RT-2/CZT in test mode to get CMOS data only  \\
\hline
Test mode, CZT + CMOS &  0xC97F    &  Set RT-2/CZT in test mode to get CZT and CMOS data only  \\
\hline
Test mode, only CZT with `n' events &  0xC9xx    &  Set RT-2/CZT in test mode to get events alone, d6-d0: no. of events/64  \\
\hline
Initialization data dump to CZT &  0xC400    &  Dumping initializing data into CZT  \\
\hline
RAM address select for RT-2/E &  0xDxxx    &  Select the 12 bit address in RAM, xxx: 12 bit address  \\
\hline
Flare decision logic command &  0xE000    &  Flare data from S `or' G  \\
\hline
 &  0xE001    &  Flare data from S `and' G  \\
\hline
RAM reset &  0xE002    &  Reset RAM  \\
\hline
PGM memory data to RAM &  0xE003    &  Get data from program memory to RAM  \\
\hline
RAM data to packets &  0xE004    &  Get data from RAM (page 62) in packets  \\
\hline
Bad to good duration set command &  0xE1xx    &  d7-d5: no. of seconds/64  \\
\hline
 &      &  d4: second select (0: interface19, 1: interface20)  \\
\hline
 &      &  d3-d0: veto\_GBLS (good, bad, light, shadow veto)  \\
\hline
RAM data write in RT-2/E (8 bit value) & 0xE2xx    & -  \\
\hline
\end{tabular}
\end{table}

\section{Flare Detection}

RT-2/E has onboard flare detection logic, which is executed every second. RT-2/E software 
switches to Solar Flare Mode (SFM) or back to Solar Quiet Mode (SQM) based on the solar mode detected 
on every 10 second boundary. The solar mode is decided based on the data from RT-2/S and RT-2/G.
There are two options by which the solar mode is decided:

\noindent
1. When RT-2/S or RT-2/G, any of them gives flare data.\\
2. When RT-2/S and RT-2/G, both give flare data.

Either of these two options can be chosen by ground command. The commands 0xE000 and 0xE001 choose 
`or' logic and `and' logic respectively. Once the solar mode is decided based on the logic, 
both RT-2/S and RT-2/G data is processed in the same mode, Solar Flare Mode (SFM, processing 
every 10 second) or Solar Quiet Mode (SQM, processing every 100 second).

Flare detection every second is done based on the accumulated values of C1, C2 and C5 counts 
(Scientific data are accumulated in 8 counters, for details, see Debnath et al. 2009) during the 
last three 100 ms intervals within the second. If the sum of the 
accumulated C1, C2 and C5 counts exceeds a set threshold value ($F_{th}$) in all these three 
intervals, then a flare is identified. These threshold values can be changed by command. Else, if 
the counts have not even exceeded the threshold at least once, then the data are assumed to be 
solar quiet data. Independent threshold values can be given for RT-2/S and RT-2/G. The commands 
for default threshold are 80FF and A0FF (see Table 25). The default value of both these thresholds 
is 0xFF, which means that the count rate should be $\ge$ $F_{th}$ x 32 i.e. 255 x 32 = 8160. 
Flare detection logic applies to RT-2/S and RT-2/G data only.


\section{Discussions and Concluding Remarks}

In a series of papers on RT-2 Experiment onboard the Coronas-Photon satellite,
we have described the technical details as well as the test and evaluation
methods. In the present paper, we have discussed the processing electronic device (RT-2/E) of RT-2.
On 30th of January, 2009, the CORONAS-PHOTON was launched successfully and all the RT-2
payload components including RT-2/E are working to our satisfaction. The Data Structure and 
Data Management softwares are working as per plan as verified by the onboard Data status. 

The onboard performance of the software was found to be very satisfactory. Since most of the data 
are slowly varying, a factor of 3 compression could be obtained, mostly with 15:1 or 14:2 options. 
In the initial days of operation, the detectors were operated without applying
the High Voltages to verify the satisfactory response to the BAD signals from the satellite. 
Once these aspects are correctly established, the detectors were operated in the SQM 
(by deliberately keeping the flare threshold high and not allowing any flare detection). It was
found that the onboard memory was adequate. The `Test Modes' were operated to calibrate the CMOS 
detectors as well as to make high time resolution observations of the Crab Nebula. In this mode, 
however, memory full signal was noticed and the system was going to the `BAD' mode. In the 
subsequent operations of the `Test modes', duration of these modes are restricted
by time-tagged commands to avoid memory overflow. Once full confidence in the overall working of 
the instrument was established, the flare detection logic was enabled with appropriate flare 
threshold commands (Instruments were operated with 0x8002 command for flare threshold value 
$F_{th}$ = 2 with count rate 64). The system was going to SFM during about 30\% of time, mostly due 
to the flare triggers encountered during high background regions. The details of the on board data 
calibration would be discussed elsewhere.

\begin{acknowledgements}
DD and TBK thank CSIR/NET scholarships and RT-2/SRF fellowship (ISRO) which supported their research 
work. The authors are thankful to scientists, engineers and technical staffs from 
TIFR/ICSP/VSSC/ISRO-HQ for various supports during RT-2 related experiments.
\end{acknowledgements}



\end{document}